\documentclass[12pt]{article} \usepackage{aaspp4,rotate} \parskip 0pt

\begin{document}
\newif\ifFigures  \Figurestrue
\def\Figure#1{\ifFigures{#1}\fi}
\def\eq#1{\begin{equation} #1 \end{equation}}
\def\E#1{\hbox{$10^{#1}$}}
\def\Ca {[25--12]}
\def\Cb {[60--25]}
\def\Cc {[100--60]}
\def\ccd{color--color diagram}
\def\ccds{color--color diagrams}
\def\about  {\hbox{$\sim$}}
\def\la     {\hbox{$\lesssim$}}
\def\ga     {\hbox{$\gtrsim$}}
\def\x      {\hbox{$\times$}}
\def\mic    {\hbox{$\mu$m}}
\def\Lo     {\hbox{$L_\odot$}}
\def\tV     {\hbox{$\tau_V$}}
\def\half   {\onehalf}
\def\threehalf {\slantfrac{3}{2}}
\def\Ref{\reference{}}
\def\HII{H{\small II}}

\def\Msg{{\em ApJ Letters}, accepted for publication March 15, 2000}
\rightline{\Msg}

\title{INFRARED CLASSIFICATION OF GALACTIC OBJECTS}

\author{\v{Z}eljko Ivezi\'c\altaffilmark{1} and
           Moshe Elitzur\altaffilmark{2}}

\altaffiltext{1} {Princeton University, Department of Astrophysical Sciences,
                  Princeton, NJ 08544--1001; ivezic@astro.Princeton.edu}
\altaffiltext{2} {Department of Physics and Astronomy, University of Kentucky,
                  Lexington, KY 40506--0055; moshe@pa.uky.edu}

\begin{abstract}

Unbiased analysis shows that IRAS data reliably differentiate between the
early and late stages of stellar evolution because objects at these stages
clearly segregate in infrared color-color diagrams. Structure in these
diagrams is primarily controlled by the density distribution of circumstellar
dust. The density profile around older objects is the steepest, declining as
$r^{-2}$, while young objects have profiles that vary as $r^{-3/2}$ and
flatter. The different density profiles reflect the different dynamics that
govern the different environments. Our analysis also shows that high mass star
formation is strongly concentrated within \about\ 5 kpc around the Galactic
center, in support of other studies.

\end{abstract}


\section{INTRODUCTION}

The IRAS all-sky survey provides a unique opportunity to classify the infrared
properties of astronomical objects from a homogeneous data set obtained with a
single facility. Cross-correlations of various catalogues with the IRAS Point
Source Catalogue (PSC) showed that certain Galactic objects tend to cluster in
well defined regions of IRAS \ccds.  Notable examples include \HII\ regions
(e.g. Hughes \& MacLeod 1989; Wood \& Churchwell 1989 [WC]) and AGB stars (van
der Veen \& Habing 1988; VH).  The reason for this clustering was not
understood, nor was it clear whether such biased analysis based on
pre-selection implies reliable selection criteria. From detailed modeling of
dusty winds we were able to validate the VH selection criterion proposed for
AGB stars and to explain its origin (Ivezi\'c \& Elitzur 1995; IE95). Here we
extend this approach to all Galactic PSC sources in an unbiased analysis of
IRAS fluxes without prior selections.

\section{IRAS DATA AND ITS CLASSES}

There are 6338 PSC sources with flux quality of at least 2 in all four IRAS
channels.  Removing the 651 sources identified as extra-galactic (Beichman et
al 1985) produces our basic data set of 5687 Galactic IR objects. For AGB stars
we have shown (IE95) that a high IRAS quality does not guarantee the flux is
intrinsic to the point source itself; high-quality 60 and 100 \mic\ PSC fluxes
can have a cirrus origin instead. Figure 1 shows that the problem afflicts all
sources, not just AGB stars.  Its x-axis is cirr3/$F_{60}$, a measure of the
ratio of cirrus background noise to the 60 \mic\ signal. Intrinsic fluxes
should have nothing to do with background emission, yet the \Cc\ color is
strongly correlated with cirr3/$F_{60}$ when this noise indicator exceeds a
certain threshold. The PSC fluxes of sources above this threshold reflect
cirrus, not intrinsic emission. Following IE95, we remove all sources with
cirr3 $> 2\,F_{60}$ to eliminate cirrus contamination. This leaves 1493
objects that can be considered a reliable representation of Galactic infrared
point sources.

We submitted these data to the program AutoClass for an unbiased search of
possible structure. AutoClass employs Bayesian probability analysis to
automatically abstract a given data base into classes (Goebel et al 1989).
AutoClass detected that the PSC sources belong to four distinct classes that
occupy separate regions in the 4-dimensional space spanned by IRAS fluxes.
Figure 2a shows the 2-dimensional projection of this space onto the \Cc--\Ca\
color plane. The IRAS colors of black body emission, which are independent of
the black body temperature as long as it exceeds \about\ 700~K, are marked by
the large dot. There is hardly any data at that point. Instead, the data are
spread far away from it, indicating that IRAS fluxes are dominated by
surrounding dust. But rather than random scatter, the data show clear
structure, with the four AutoClass classes occupying well defined color
regions. As a further check on the reality of these classes we constructed
their Galactic distributions, shown in Figure 2b.

Classes A and B clearly separate in the \ccd. Classes C and D, on the other
hand, are mixed together and are distinguished by their flux levels --- class C
fluxes are typically 10--100 times higher in all 4 bands. In principle, a
single family could produce such behavior if split into nearby and distant
objects, but this is not the case here. Extragalactic and heliocentric
selection effects are ruled out by the Galactic distributions. Both classes
are comprised of Galactic disk objects, but the median flux of class C is 16
times higher.  If the two were drawn from the same population, class C sources
would be on the average 4 times closer and their Galactic latitude
distribution 4 times wider.  Instead, the latitude histograms are essentially
the same, both are centered on $b = 0\arcdeg $ with full-width at half-maximum
$1.5\arcdeg \pm 0.3\arcdeg$ for class C and $1.5\arcdeg \pm 0.1\arcdeg$ for
class D. Class D fluxes vary by more than \E3, consistent with a population
distributed throughout the entire Galactic plane.\footnote{Class A fluxes have
a similar dynamic range.} In contrast, class C fluxes vary by a factor of only
\about\ 20 and exceed \E3 Jy at both 60 and 100 \mic\ for all sources. This is
consistent only with a population confined to, at most, \about\ 5 kpc around
the Galactic center with a minimal luminosity of \E3 \Lo. The Galactic
longitude distributions corroborate these conclusions. Although both class C
and D sources channel most of their luminosity to the far-IR, their spatial
and luminosity distributions are very different.

\section{THEORY}

In IE95 we identified the cause of the particular IRAS colors of AGB stars: The
dusty wind problem possesses general scaling properties. For a given dust
composition, the solutions are predominantly determined by a single input
parameter --- the overall optical depth.  All other input is largely
irrelevant. As a result, the solutions occupy well defined regions in \ccds.
Indeed, the IRAS measured colors of AGB stars fall in the regions outlined by
wind solutions with visual optical depth \tV\ \la\ 100.

%

In Ivezi\'c \& Elitzur (1997; IE97) we showed that scaling is a general
property of radiatively heated dust under all circumstances, not just in AGB
winds.  The most general dust radiative transfer problem contains only two
input quantities whose magnitudes matter --- \tV\ and the dust temperature
$T_1$ at some point. All other input is defined by dimensionless, normalized
profiles that describe (1) the spectral shape of the external radiation, (2)
the spectral shape of the dust absorption and scattering coefficients, and (3)
the dust spatial distribution. Physical dimensions such as, e.g., luminosity
and linear sizes are irrelevant. Scaling applies to arbitrary geometries, and
in IE97 we conducted extensive numerical studies of its consequences for
spherical shells. With a black body spectral shape for the heating radiation,
the temperature $T$ hardly affects IRAS colors when $T >$ 2000~K. The reason
is that $T$ is much higher than the Planck equivalent of the shortest IRAS
wavelength (12 \mic), which is only 240 K. Although this result was obtained
in spherical solutions, its validity is general. Since neither its luminosity
nor spectral shape matter, the heating source is quite irrelevant for the IRAS
colors. Different objects segregate in IRAS \ccds\ not because they have
different central sources, but rather because their dust shells are different.

In our extensive study of spherical shells we examined also the effect of
$T_1$, which was selected as the dust temperature on the shell inner boundary.
As with the heating radiation, and for the same reasons, $T_1$ hardly affects
IRAS colors when varied over the plausible range of dust sublimation
temperatures 700--2000 K (IE97). This leaves the dust's composition, optical
depth and spatial distribution as the only potentially significant reasons for
the variation of IRAS colors among different families of Galactic objects.
However, dust properties generally do not show large variations among
different objects and thus cannot be expected to induce substantial variations
in infrared emission. The same applies to \tV, whose range of values is similar
in most families. We conclude that {\em objects segregate in IRAS \ccds\
primarily because their spatial dust distributions are different.}

%

We corroborate these conclusions in detailed modeling with the code DUSTY
(Ivezi\'c, Nenkova \& Elitzur 1999).  In these calculations, a point source
emitting as a black body with temperature 5000~K is surrounded by dust whose
absorption and scattering coefficients are those of standard interstellar mix.
The dust is distributed in a spherical shell whose temperature on the inner
boundary is $T_1$ = 1000~K.  The shell density profile is taken as a power law
$r^{-p}$, with $p$ a free parameter. Scaling ensures that the solution set for
each $p$ is a one-parameter family characterized by \tV, leading to the
distinct tracks shown in the \Cb--\Ca\ \ccd\ in figure 3. These tracks
properly delineate the distribution of IRAS sources and we have verified that
the same applies also to the \Cc\ colors shown in fig.\ 2. Tracks for
different $p$ differ in the color-color regions that they cross and in the
distance induced by \tV\ variation; the steeper the density profile, the
smaller is the distance along the track for the same change in \tV. Both
properties arise from the differences in relative amounts of material placed
at different temperatures, and both are reflected in the data. Class A sources
are well explained by the $p$ = 2 track, indeed they cluster close to the
black-body point.  In contrast, classes C and D are explained by tracks of
much flatter density distributions and are located rather far from the
black-body colors; it takes only \tV\ \about\ 0.1 to move an object along
these tracks all the way from the black body point to the region populated by
IRAS sources.

Class B cannot be reasonably explained by the same models, even though its
colors can be produced by extending the $p$ = 2 track beyond \tV\ = 100. In
addition to the implausible optical depths this requires, class B is unlikely
to be the high-\tV\ end of class A given their different Galactic
distributions. An alternative to \tV-variation is to modify colors by removal
of hot dust. The displayed tracks have $T_1$ = 1000 K, as appropriate for
shells whose inner boundaries are controlled by dust sublimation, but in
detached shells $T_1$ is both lower and arbitrary. The dashed-line tracks in
figure 3 show the effect of lowering $T_1$ on the $p$ = 2 track. The
difference between $T_1$ = 1000 K and 300 K is marginal because both are
higher than the Planck equivalent of 12 \mic. However, further reduction in
$T_1$ alters the track significantly, adequately explaining class B colors.

\section{DISCUSSION}

Our modeling confirms that the primary reason for different IRAS classes is
different dust density distributions.  Only 5\% of the sources require an
additional variation of the dust temperature on the shell inner boundary.
Spherical shells were employed here as the simplest method to model extended,
three dimensional dust distributions. Power laws were used just for
illustration purposes and although their actual values should not be taken
literally, they provide a good indication of the overall behavior of the
density distributions in the four IRAS classes. The presence of disks,
expected in various sources, should not significantly affect our conclusions.
The standard disk spectrum $F_\nu \propto \nu^{1/3}$ produces a single point
in \ccds\ and thus cannot effect the observed scatter. This spectrum is
modified if the disk is embedded in an extended shell, but then the disk is
expected to dominate only at sub-mm and mm wavelengths, longer than those
observed by IRAS (Miroshnichenko et al 1999). Flared disk emission can be
shown equivalent to that from a flat disk embedded in an appropriate spherical
shell.  Therefore, such a configuration, too, cannot modify our conclusions.

We have queried the SIMBAD database about our sample sources. SIMBAD
identifications are occasionally ambiguous (``maser"), sometimes less than
informative (``part of cloud"), and reliability is not always certain.
Nevertheless, they provide useful clues when there are major trends in the
data. Class A returned the most decisive results --- 88\% of its members have
possible optical identifications, of which roughly 90\% are commensurate with
AGB stars. Since class A obeys the VH criterion, this corroborates our earlier
finding (IE95) that this criterion is both sufficient and necessary for AGB
selection. In IE95 we present a thorough analysis of the IRAS fluxes of AGB
stars, including color tracks geared specifically for these objects (single
chemistry grains, appropriate stellar temperature, etc.). While that analysis
was necessary to verify the VH criterion, the $p$ = 2 track presented here
captures the essence of the more detailed study, demonstrating that the
density distribution is the leading factor in controlling IRAS colors. These
circumstellar shells have the steepest density distribution, setting them
apart in the \ccds. In addition, the color tracks are primarily controlled by a
single parameter, \tV, hence the compactness of this class color region.

Class B had the same identification rate but its composition not as
homogeneous. Planetary nebulae comprise 40\% of positive and possible
identifications. At 13\%, the only other significant group is ``emission-line
stars", a classification consistent with planetary and reflection nebulae.  The
remaining identifications span a variety of objects, indicating that detached
shells may occasionally form under different circumstances. Significantly, the
optical depths required for class B are lower than for the others, the
dashed-line tracks in fig.\ 3 terminate at \tV\ = 10.

SIMBAD identification rates for the two other classes are much lower --- only
32\% for class C and 21\% for D.  Among the 38 class C identifications, \HII\
regions comprise the only significant group with 15 (40\%).  In class D, too,
\HII\ regions comprise the single largest group with 23\% of the
identifications, followed by young stellar objects at 11\% and planetary
nebulae at 10\%. These two classes are clearly dominated by star formation and
early stages of stellar evolution, in agreement with their Galactic
distributions and with previous attempts to associate star-forming regions
with IRAS colors. In the most extensive study of this kind, 83\% of 1302 IRAS
selected sources were found to be embedded in molecular clouds and thus trace
star formation, and the selection criterion (\Ca\ $\ge$ 0, \Cb\ $\ge$ 0.4)
essentially identifies classes C and D (Wouterloot \& Brand 1989; Wouterloot
et al 1990).

The IRAS colors of classes C and D imply dust density distributions flatter
than for AGB stars.  These colors are spread over large regions, reflecting
variation in density profiles in addition to optical depth. The spread in \Cc\
colors is smaller than in \Cb\ because all shells are optically thin at both
100 and 60 \mic\ while their 25 \mic\ optical depth can become significant
(IE97). Class C colors occupy a sub-region of class D and are produced by the
optically-thick end (\tV\ \ga\ 1) of flat density distributions.  Objects whose
colors fall in that region can belong to either class C or D.  However, among
all sources with class D colors, those with high fluxes ($>$ 1000 Jy at both
60 and 100 \mic) concentrate in a compact color region, hence the separate
class C. Since all class C sources have $L >$ \E3 \Lo, they are high-mass
objects and their concentration in the inner \about\ 5 kpc of the Galaxy shows
that the high-mass star formation rate decreases with distance from the
Galactic center. This result is in agreement with studies of the initial mass
function inside and outside the solar circle (Garmany et al 1982; Wouterloot et
al 1995; Casassus et al 1999).

Our results explain the findings of all earlier studies that were based on
object pre-selection, and reveal the limitations of that approach. By example,
consider the WC study. After identifying IRAS counterparts of known
ultracompact \HII\ regions, WC proposed the corresponding colors as a necessary
and sufficient selection for all ultracompact \HII\ regions and proceeded to
estimate the birthrate of O stars. Codella et al (1994) then found that most
\HII\ regions in a more extended catalog indeed obeyed the WC color criterion.
However, that included also diffuse, not just compact \HII\ regions, therefore
WC overestimated the O star birthrate. This clearly demonstrates the
shortcomings of any classification based on a pre-selected population.  The
unbiased analysis presented here shows that IRAS colors reflect primarily the
dust density profiles of circumstellar shells and provide a unique indication
of the underlying object only in the case of AGB stars. IRAS data in itself is
sufficient for differentiating young and old stellar objects; apart from a
limited number of detached shells, IRAS sources belong to two distinct groups
as is evident from both the \ccds\ and the Galactic distributions: (1) class A
sources are at the late stages of stellar evolution and (2) class C and D
sources are objects at the early evolutionary stages. This differentiation
occurs because the density distributions of dust around young stellar objects
have flatter profiles, reflecting the different dynamics that govern the
different environments.

\acknowledgments

Support by NASA and NSF is gratefully acknowledged.

%
%

\newpage

\figcaption[fig1.ps]{\Cc\ color vs.\ cirr3/$F_{60}$, where $F_{60}$ is the 60
\mic\ IRAS flux and cirr3 is the IRAS surface brightness at 100 \mic\ around
the point source.  The flux quality for all sources is at least 2 in all four
IRAS bands, and 3 for those marked with dark dots. In spite of the high flux
quality, the strong correlation with the cirrus emission at large values of
cirr3/$F_{60}$ is evident.}
\bigskip

\figcaption[fig2.ps]{(a) \Cc--\Ca\ \ccd\ for all the Galactic IRAS sources free
from cirrus contamination. The four classes identified by AutoClass are marked
with different symbols: Class A (282 sources) --- crosses, B (80 sources) ---
circles, C (119 sources) --- squares, D (1013 sources) --- dots. The large dot
marked BB denotes the colors of black body emission with temperature $T > 700$
K. (b) Galactic distributions of the four classes.}
\bigskip

\figcaption[fig3.ps]{\Cb--\Ca\ \ccd. Data are for the same sources presented
in figure 2 using the same symbols. The full lines are model tracks for
spherical dust shells with $T_1$ = 1000~K on their inner boundary and $r^{-p}$
density profiles with $p$ = 0, \half, 1, \threehalf\ and 2, as marked.
Position along each track increases with dust optical depth \tV\ away from the
common origin \tV\ = 0 at the black-body colors, with \tV\ = 0.1 marked by
solid triangle and \tV\ = 10 by solid square.  The endpoint for each track is
\tV\ = 100. Dashed-line tracks show the effect on the $p$ = 2 track of lowering
$T_1$ to 300 K (lowermost track), 200 K, 150 K and 100 K (uppermost track).
These tracks end at \tV\ = 10.}
\bigskip


\Figure{\bigskip \centerline{\epsfclipon \epsfxsize=\hsize \epsfbox{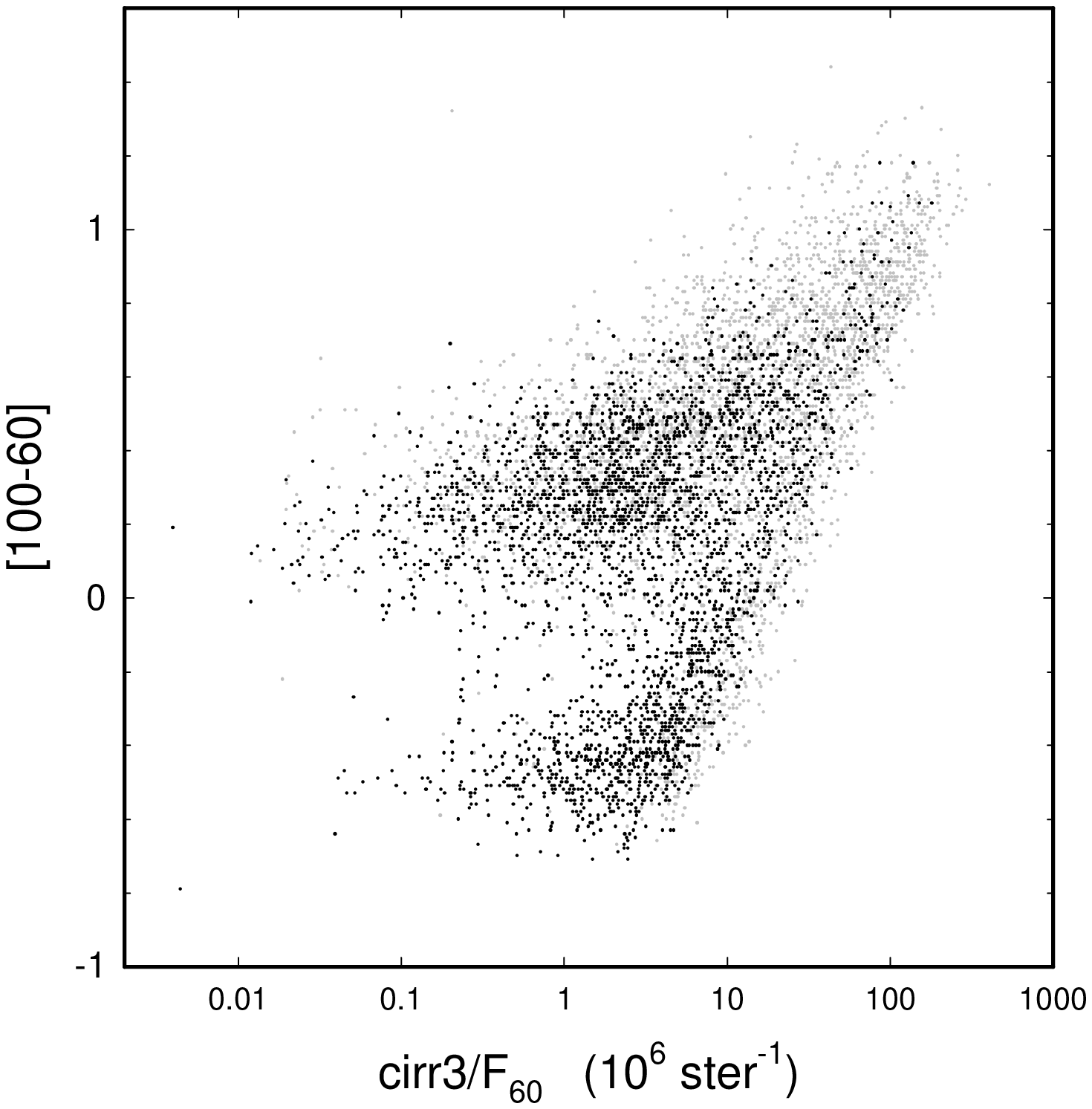}}}

\Figure{\bigskip \epsfclipon \epsfxsize=0.9\hsize \epsfbox{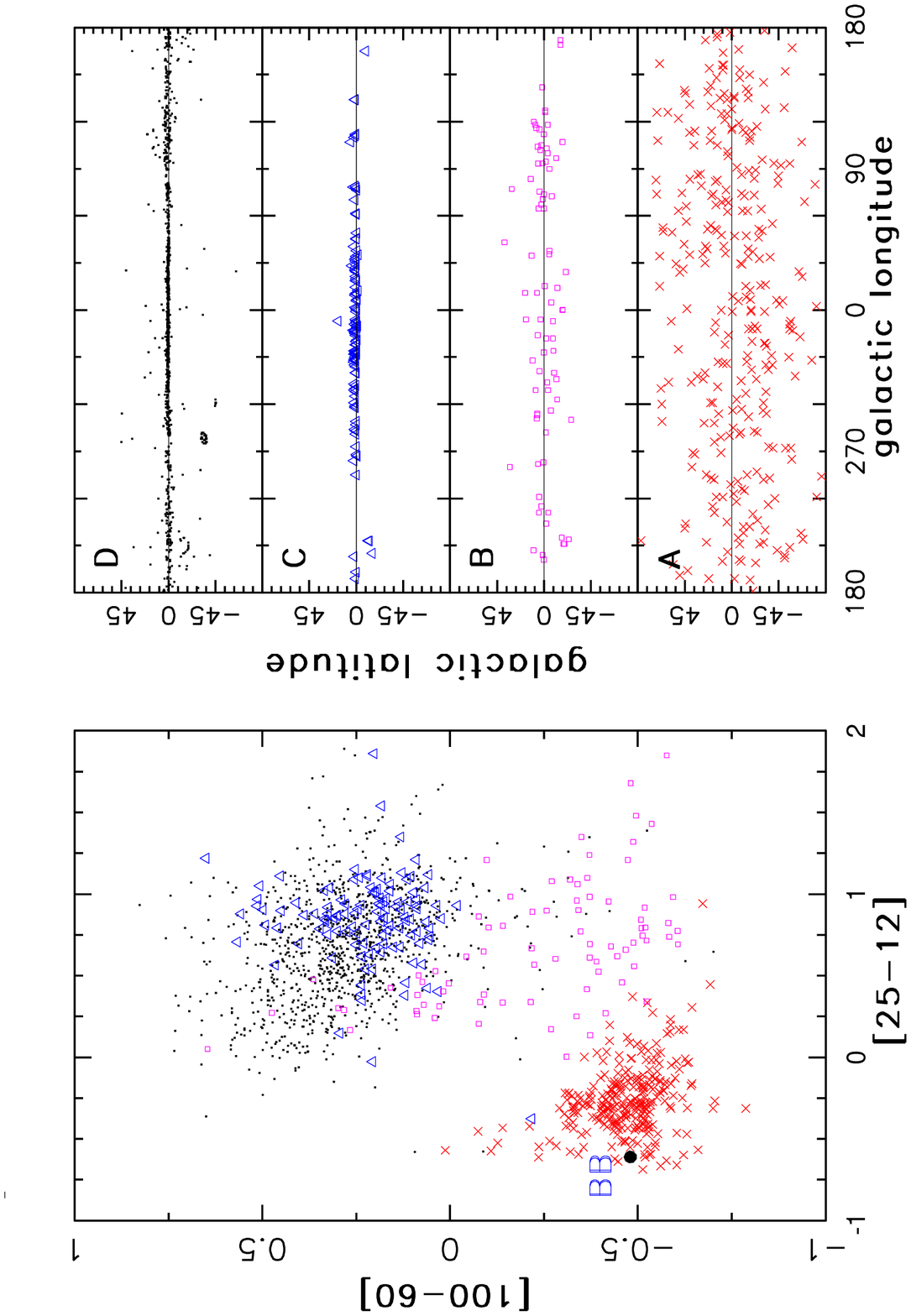}}

\Figure{\bigskip \centerline{\epsfclipon \epsfxsize=\hsize \epsfbox{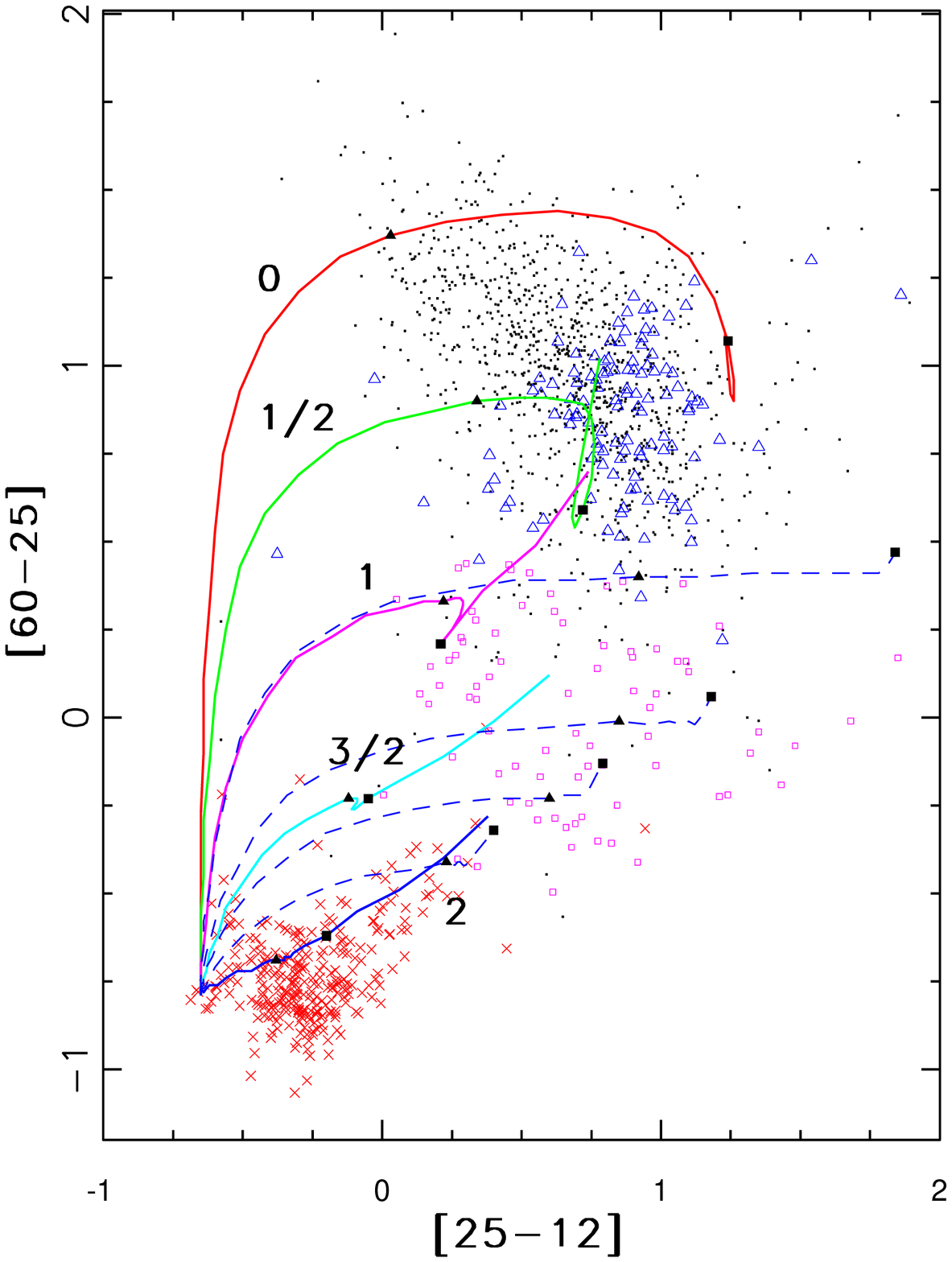}}}


\begin{references}

\Ref Beichman, C.A., Neugebauer, G., Habing, H.J., Clegg, P.E.
     \& Chester, T.J., 1985, {\em IRAS Catalogs and Atlases}
     (US GPO, Washington, DC)
\Ref Casassus, S., Bronfman, L., May, J. \& Nyman, L-\AA, 1999, A\&A submitted
     (astro-ph/9912340)
\Ref Codella, C., Felli, M. \& Natale, V., 1994, A\&A, 284, 233
\Ref Goebel, J., et al 1989, A\&A, 222, L5
\Ref Garmany, C.D., Conti, P.S. \& Chiosi, C., 1982, ApJ, 263, 777
\Ref Hughes, V. A. \& MacLeod, G. C. 1989, AJ, 97, 786
\Ref Ivezi\'c, \v Z., Elitzur M., 1995, ApJ, 445, 415 (IE95)
\Ref Ivezi\'c, \v Z., Elitzur M., 1997, MNRAS, 287, 799 (IE97)
\Ref Ivezi\'c, \v Z., Nenkova, M., Elitzur, M., 1999, User Manual for DUSTY,
     University of Kentucky Internal Report, accessible at
     http://www.pa.uky.edu/$\sim$moshe/dusty
\Ref Miroshnichenko, A., Ivezi\'c, \v Z., Vinkovi\'c, D. \& Elitzur M., 1999,
     ApJ, 520, L115
\Ref van der Veen, W.E.C.J., \& Habing, H.J. 1988, A\&A, 194, 125 (VH)
\Ref Wood, D. O. S. \& Churchwell, E. 1989, ApJ, 340, 265 (WC)
\Ref Wouterloot, J.G.A. \& Brand, J., 1989, A\&AS, 80, 149
\Ref Wouterloot, J.G.A., Brand, J., Burton, W.B. \& Kwee, K.K., 1990,
     A\&A, 230, 21
\Ref Wouterloot, J.G.A., Fiegle, K., Brand, J. \& Winnewisser, G., 1995,
     A\&A, 301, 236

\end{references}
\end{document}